\documentclass[10pt,a4paper]{article} 
\usepackage{jheppub}
\usepackage[utf8]{inputenc}
\usepackage[english]{babel}
\usepackage{graphicx}
\usepackage{amsmath}
\usepackage{amsfonts}
\usepackage{amssymb}
\usepackage{hyperref}
\usepackage{url}

\usepackage{hyperref}
\usepackage{url}

\newcommand{\bb}[1]{\boldsymbol{#1}}
\title{Multi-channel pattern reconstruction through $L$-directional associative memories}

\author[a,1]{Elena Agliari,}
\author[b,1]{Andrea Alessandrelli,}
\author[a]{Paulo Duarte Mour\~ao,}
\author[a,1]{Alberto Fachechi}
\affiliation[a]{Department of Mathematics,
Sapienza Università di Roma, Piazzale Aldo Moro 5, 00185 Rome, Italy}
\affiliation[b]{Department of Informatics, Università di Pisa, Largo Bruno Pontecorvo 3, 56127 Pisa, Italy}
\affiliation[1]{Gruppo Nazionale per la Fisica Matematica (GNFM-INdAM), Italy}
\emailAdd{alberto.fachechi@uniroma1.it}
\abstract{We consider $L$-directional associative memories, composed of $L$ Hopfield networks, displaying imitative Hebbian intra-network interactions and anti-imitative Hebbian inter-network interactions, where couplings are built over a set of hidden binary patterns. We evaluate the model's performance in reconstructing the whole set of hidden binary patterns when provided with mixtures of noisy versions of these patterns. Our numerical results demonstrate the model's high effectiveness in the reconstruction task for structureless and structured datasets.}

\begin{document}

\maketitle


\section{Introduction and related works}
The Hopfield model \citep{hopfield1982neural} is a cornerstone in the investigation of artificial neural networks, the main reason for such an importance lying in the crucial intuition that functionalities of artificial neural networks can be framed, from a physical point of view, as emerging collective properties much as like the thermodynamic properties of particle systems. Since its introduction, and especially after the solution by Amit, Gutfreund and Sompolinsky \citep{amit1987statistical}, the Hopfield model -- and related models of associative memory -- has attracted a continuously growing attention and today we have a clear picture of its working principles, including issues that may impair its pattern-reconstruction functionalities.
Among these, spurious attractors have been examined in detail and several modifications have been proposed in order to reduce their attractiveness, retaining the pairwise interaction structure between the units (e.g., \cite{dotsenko1991statistical,fachechi2019dreaming}) or extending the interaction order as in the dense associative memories (e.g., \cite{krotov2016dense}). 

Remarkably, in recent years, pattern reconstruction and variations on the theme of the Hopfield model have gained broad significance and found applications in various fields. For instance, from a purely numerical perspective, they have been employed in matrix (and possibly tensor) factorization through decimation schemes (see, for example, \citep{camilli2023matrix} and references therein). Further, autonomous pattern reconstruction has today become one of the key aspects in modern Machine Learning theory, as it allows to shed light on the ability of neural networks to extract patterns from set of data and enable feature learning \citep{bengio2012unsupervised,aiudi2025local}, as well as investigating generalization in simplified settings \citep{negri2023storage, negri, agliari2024regularization}.  

In this work, we explore the possibility to reconstruct binary hidden patterns by means of $L$-directional associative memories, assuming that the Hebb coupling matrix built on these patterns is given, along with additional information in terms of mixtures of corrupted versions of the same hidden patterns. We present numerical results across various settings, demonstrating strong performance for both structureless and structured datasets.

\section{The model: $L$-directional associative memory}
The $L$-directional generalization of the Hopfield model proposed in  \cite{agliari2025networks} is an energy-based model made up of an assembly of $L$ Hopfield networks, each referred to as a layer, whose neuronal configurations are denoted as $\boldsymbol \sigma^a \in \{-1, +1\}^N$ with $a=1, ...,L$. The model exhibits both intra- and inter-layer interactions. Specifically, given a realization of patterns $\bb \xi^\mu \in \{-1, +1 \}^N$, with $\mu=1,...,K$, 
%
%
the energy function reads as $E=-N \sum_{a,b=1}^L g_{a,b} m_\mu ^a m_\mu ^b$, where $m _\mu ^a= N^{-1}\sum_{i=1}^N \xi^\mu_i \sigma_i ^a$ is the overlap between the $a$-th layer configuration and the $\mu$-th pattern, while $g_{a,b}$ is chosen in such a way that $g_{a,a}=1$ -- hence reproducing the usual Hopfield energy function within each layer -- and $g_{a,b}=-\lambda$ for $a\neq b$, with $\lambda \in \mathbb R_+$ being a tunable hyper-parameter -- hence discouraging the retrieval of the same pattern by different layers. As shown in \cite{agliari2025networks} focusing on the case $L=3$, this network is able to disentangle mixtures of patterns, like the notorious spurious states $\boldsymbol x= \textrm{sgn}(\sum_{\nu=1}^L \boldsymbol \xi^{\nu})$, in a wide region of the parameter space, that is, by supplying $\boldsymbol x$ as input configuration on each layer, the system can relax to the target configuration $(\boldsymbol \sigma^1,\bb \sigma^2, \boldsymbol \sigma^3) = (\bb \xi^1, \bb\xi^2,\bb\xi^3)$, or any suitable permutation that ensures the retrieval of each single pattern in the original mixture \footnote{The scheme here adopted can be interpreted as a parametric algorithm to achieve Independent Component Analysis (ICA) where data are available in a random feature setting \citep{negri2023storage,negri}. Notice, however, that the proposed scheme only gives the source vectors (the hidden patterns) involved in the mixture combinations but not the associated coefficients, whose determination requires additional procedures.}. However, it was also noticed that the energy function is invariant under a global spin-flip of all layers, but it is not invariant if layer configurations are reversed individually, namely $\bb\sigma^a \to -\bb\sigma^a$ for some $a=1,\dots,L$. As a consequence, beyond the target configuration $(\boldsymbol \sigma^1,\bb \sigma^2, ..., \boldsymbol \sigma^L) = (\bb \xi^1, \bb\xi^2,\dots,\bb\xi^L)$, also configurations such as $(\bb \xi^1, \dots,-\bb\xi^1,\dots,\bb\xi^1)$ can exhibit strong attractive power for the neural dynamics, thus impairing the disentangling capabilities of the model. One way to prevent these undesired attractors and reduce their attraction basins, is to break the quadratic nature of the energy function by considering the square of inter-layer contributions in the energy function. Also, an external field $\bb h ^a$ (modulated by a field strength $H$) driving the dynamics during evolution can be applied on each layer. 
Putting all pieces together and denoting with $\bb\sigma$ the overall configuration of the composite network, the resulting energy function reads:
\begin{equation}\label{eq:energy}
    E_{N,\bb\xi} (\bb\sigma)=- N \sum_{a=1}^L\sum_{\mu=1}^K (m_\mu^a)^2+N\lambda \sum_{a \neq b=1}^L(\sum_{\mu=1}^Km_\mu^a m_\mu^b)^2-  H\sum_{a=1}^L\sum_{i=1}^N h^a_i\sigma^a_i.
\end{equation}
This energy function results in a larger portion of the parameter space where the system successfully disentangle spurious states \citep{agliari2025networks}. In the present paper, we show that this model can be employed even for more challenging tasks, as detailed in the following section. Before proceeding, we explicit the neuronal dynamics applied to the system: allowing for the presence of stochastic noise, tuned by the thermal parameter $\beta\in \mathbb R_+$, the neuronal configuration is synchronously updated as
\begin{equation}\label{eq:update}
    \bb \sigma^a (t+1)= \text{sgn}[\tanh(\beta\tilde{\bb  h}^a (t))+\bb u^a (t)],
\end{equation}
with $t$ being the discrete time, $\bb u^a (t)\sim\mathcal U ([-1,+1]^N)$ i.i.d. providing the source of noise, and $\tilde h_i ^a$ being the net field acting on the spin $i$ in the $a$-th layer. This can be expressed as
\begin{equation}\label{eq:net_field}
    \tilde {\bb h} ^a (t) =\bb h^{(a\to a)}(t)+ \sum_{b\neq a}\bb h^{(b\to a)}(t)+H \bb h^a.
\end{equation}
where, denoting with $\bb J =N^{-1} \bb \xi \bb \xi^T$ the Hebbian matrix, $ \bb h^{(a\to a)}(t) = \bb J\cdot \bb \sigma ^a(t) $ and $\bb h^{(b\to a)}=-\lambda \bb h^{(b\to b)} (t)  (\bb \sigma^b (t) \cdot \bb h^{(a\to a)}(t)) $ are, respectively, the intra- and inter-layer internal fields at time $t$, acting on the layer $a$. 

\section{Tasks and results: multi-channel pattern reconstruction}
Given the ability of the model (\ref{eq:energy}) to disentangle spurious states, it is worth investigating whether it can reconstruct patterns also from more general combinations. Specifically, we provide the model with a fixed number $m$ of inputs of the form $\bb x^\gamma =\text{sgn} (\sum_{\mu =1}^K c_{\mu}^\gamma\bb\xi^\mu)$, with $c_{\mu}^\gamma$ for $\mu=1, ...,K$ and $\gamma = 1, ...,m$ to be particularized according to the setting
\footnote{The application $\xi^\mu_i \to x_i ^\gamma = \text{sgn} (\sum_{\mu =1}^K c_{\mu}^\gamma\xi^\mu_i)$ can be interpreted as a (non-linear) random mapping of the $K$-dimensional vectors $\bb \xi_i$ onto a space with dimension $m$, or, equivalently, as the response of a perceptron with $K$ inputs and $m$ outputs, with the spin index $i$ labeling data points.}.
Next, we run the dynamics (\ref{eq:update}) and check whether the final configuration\footnote{This is reached after a time $t$ long enough to ensure the stationarity of the temporal average of the overlaps $m_\mu ^a$ over a sufficiently wide time window.} $\bar {\boldsymbol \sigma} = \{ \bar {\boldsymbol \sigma}^1, ..., \bar {\boldsymbol \sigma}^L\}$ has reached the target configuration $(\boldsymbol \xi^1, ..., \boldsymbol \xi^L)$, or any proper permutation. 
We emphasize that, in fact, there is no guarantee that the system relaxes to a disentangled representation of the inputs; thus, we should include specific quality checks for candidate reconstructed patterns. Remarkably, since the patterns $\{\boldsymbol \xi \}_{\mu=1}^K$ are not available, a direct comparison between $\bar {\boldsymbol \sigma}$ and ${\boldsymbol \xi}^{\mu}$ is not feasible and, as explained in the following, these checks leverage the algebraic properties of a suitable transformation of $\bb J$. 

Let us start with the following setting: assume that the ground patterns are Rademacher, namely each entry is extracted as $\mathcal  P(\xi^\mu_i =\pm 1)=1/2$ for all $i=1,\dots,N$ and $\mu=1,\dots,K$, and hidden, while we have access to the mixtures $\boldsymbol x^{\gamma}$, $\gamma =1, ...,m$ as defined above with $c_{\mu}^\gamma\sim\mathcal N(0,1)$ i.i.d. for $\mu=1, ...,K$ and $\gamma = 1, ...,m$. For each combination $\gamma$, we set $\bb h^a= \bb x^\gamma$ for all $a=1,\dots,L$ and let the system evolve under neural dynamics (\ref{eq:update}), whence we collect the $L\cdot m$ final configurations $\{\bar {\bb \sigma}^l\}_{l=1}^{Lm}$ as candidate reconstructed pattern; clearly, if we want to recover the whole set of hidden patterns we need $Lm \ge K$. At this point, we notice that: $i)$ there could be duplicate candidates, {\it i.e.} configurations in $\{\bar {\bb \sigma}^l\}_{l=1}^{Lm}$ with high mutual overlap, and $ii)$ configurations stacked in some spurious state. To address point $i)$, we compute the mutual overlap $q_{lk} = N^{-1}\sum_{i=1}^N \bar\sigma_i ^l \bar \sigma_i ^k$, and discard duplicates if $q_{lk}>0.5$ (a sufficiently high threshold for the random pattern setting). Regarding the point $ii)$, we recall that the true patterns $\bb\xi^\mu$ are eigenvectors (with a degenerate eigenvalue 1) of the pseudo-inverse coupling matrix $J_{ij}^{K} = N^{-1}\sum_{i,j}^N \sum_{\mu,\nu=1}^L \xi^\mu_i C_{\mu,\nu}^{-1}\xi^\nu_j$, with $C_{\mu,\nu} = N^{-1} \sum_{i=1}^N \xi^\mu_i \xi^\nu_i$ being the pattern correlation matrix \citep{kohonen1973representation,personnaz1985information,kanter1987associative}. We can obtain the latter coupling matrix as fixed point of the iterative algorithm \citep{fachechi2019dreaming}
$$
\bb J_{k+1}=\bb J_k+ \frac{\epsilon }{1+\epsilon k} (\bb J_{k}-\bb J_{k}^2),
$$
with $\epsilon< ({\lVert \bb C\lVert}-1)^{-1}$ being the unlearning strength and the initial condition being Hebb's matrix: $\bb J_0=\bb J$. Thus, in order to solve $ii)$ and discard spurious states, we require $\bar{\bb\sigma}^l \bb J ^{K} \bar {\bb \sigma}^l/N >0.8$.

\begin{figure}
    \centering
    \includegraphics[width=0.85\textwidth]{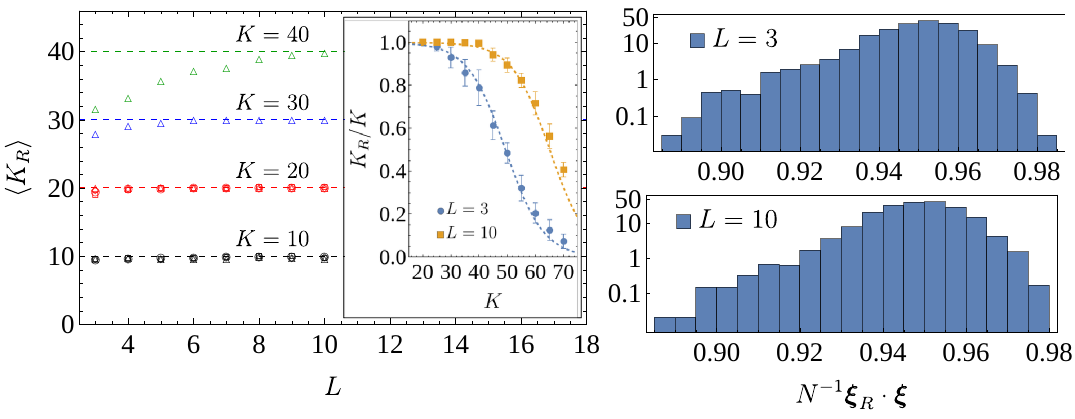}
    \caption{Summary of results for pattern reconstruction by general combinations $\textrm{sgn}(\boldsymbol c^{\gamma} \boldsymbol \xi)$. In the left plot, we present the average number of reconstructed patterns as a function of $L$ for various values of $K$. For $K=10,20$, we reported the results starting with $m=10,20,30,40,50$ combinations shown by different symbols (as they lead to the same values of $K_R$ symbols are collapsed), while for $K\ge 30$ only the results for $m=50$ are shown. In the inset of the same plot, we reported the fraction of reconstructed patterns as a function of $K$ for $L=3$ (low-complexity machine) and $L=10$ (high-complexity scenario). The dashed lines represents a fit of the form $K_R = K/[1+\exp(\frac1\kappa(K_R-K_c))]$. In particular, for $L=3$ we have $K_c\approx 50$, while for $L=10$ the critical number of patterns is $K_c \approx 65$. The numerical results are averaged over 10 different realizations of the patterns $\bb \xi^\mu$ and the matrix $\bb c$. In the right plots, we present the aggregated results for the overlap between reconstructed patterns and the hidden ones: the histograms are realized by collecting all the results with fixed $L=3$ and $L=10$ (that is, for all the values of $K$ and $m$). The network size is fixed to $N=2000$, while $\beta=2$, $\lambda=0.2$, $H=0.1$.}
    \label{fig:results_1}
\end{figure}

Out of the $mL$ collected final configurations, we now select those that fulfill the last inequality and are distinct as prescribed in point $i)$. The items of this subset are denoted as $\bb \xi_R^{\ell}$, $\ell =1, ..., K_R$ to emphasize that they provide a reconstruction of the hidden patterns; the cardinality $K_R$ represents the number of the reconstructed hidden patterns. We stress that this outcome is reached by simply exploiting the knowledge of the Hebbian matrix and the set of $m$ mixtures. Finally, to assess the quality of the reconstruction achieved by $\bb \xi_R^\ell $ we compute the quantity $m_{\ell} = \max_{\nu} [N^{-1} \bb \xi_R^\ell \cdot \bb \xi^\nu]$. Based on this procedure, we performed extensive Monte Carlo simulations and evaluated the expectation of $K_R$ and the quality of reconstruction $N^{-1} \boldsymbol \xi_R\cdot \boldsymbol \xi$. The results of the algorithm described here are presented in Fig. \ref{fig:results_1}.
In the left plot, we report the average number $K_R$ of reconstructed patterns as a function of the number of channels $L$ for various values of $K$; clearly, the higher the complexity of the machine, the more effective the pattern extraction. In particular, as the number of patterns 
$K$ to be extracted increases, the complexity required to successfully accomplish the task also rises. This is evident from the inset of the same plot, reporting the fractions of reconstructed patterns as a function of $K$ for $L=3,10$. In any case, the individual quality of the reconstructed patterns is high and slightly improves by increasing $L$, as shown by the histograms on the right.

\begin{figure}[h!]
    \centering
    \includegraphics[width=0.85\textwidth]{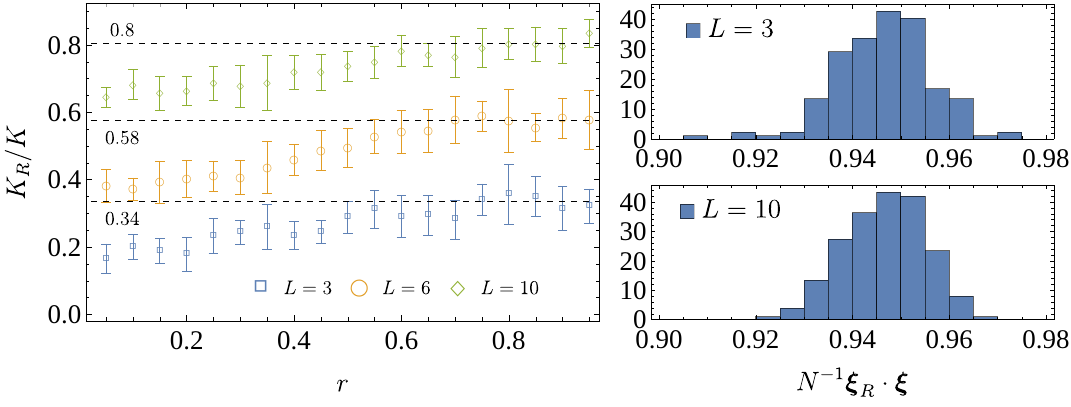}
    \caption{Summary of the results for pattern reconstruction with unsupervised combinations of examples. The left plot shows the dependence on the dataset quality $r$ of the fraction of reconstructed patterns (here, $K=50$) for different complexity of the machines: $L=3,6,10$. The horizontal dashed lines stand for the asymptotic values of $K_R/K$ at $r=1$. The results are averaged over 10 different realizations of the patterns and the associated dataset. On the right side, we reported the histograms of the overlap of reconstructed patterns with the true ones. The combinatinations of examples are $m=50$, the number of training examples (the mini-batches used to generate them) is fixed to $n=25$, the number of examples per class is $M=500$. The network size is $N=2000$, while $\beta =2$, $\lambda=0.2$, $H=0.1$.}
    \label{fig:results_2}
\end{figure}

In the second setting we address a more realistic situation, where the accessible mixtures of hidden patterns are replaced by mixtures of noisy versions of the hidden patterns, referred to as examples. These are denoted as $\{\bb \xi^{\mu,A}\}_{\mu,A=1}^{K,M}$, with $\mu$ labeling the class and $A$ distinguishing different items associated to the same pattern. Moreover, in the unsupervised scenario there is no {\it a priori} distinction of the examples in classes, that is, the label $\mu$ is unknown. To mimic this setting, we produce a synthetic dataset in the following way: first, extract the (hidden) patterns $\bb \xi^\mu$ as before, then we generate the examples by applying a multiplicative Bernoulli noise with quality parameter $r\in (0,1)$, specifically $\xi^{\mu,A}_i = \chi^{\mu, A}_i \xi^\mu_i$, with $\mathcal P(\chi ^{\mu,A}_i = \pm1 )= \frac{1\pm r}{2}$.\footnote{The role of the parameter $r$ as the quality of the dataset is clear since, for $r=1$, the examples are perfect copies of the hidden pattern, while for $r=0$ examples are just random vectors carrying no information about the hidden patterns.} Taking a mini-batch of size $n$ at random from the dataset, we can construct combinations of the form $x _i = \text{sgn}(\sum_{p=1}^n \xi_i^{\mu_p,A_p})$ mixing examples in different classes (thus, in this setting, the coefficients $c^\gamma_{\mu,A}$ are 1 if the corresponding item lies in the mini-batch, 0 otherwise). For large enough $n$, we would also have a large number of examples belonging to the same class, so that (denoting with $n_\mu$ the multinomial random variable representing the number of examples belonging to the class $\mu$ in a specific mini-batch) by virtue of the central limit theorem $\sum_{p=1}^{n_\mu}\xi^{\mu,A_p}=\xi^\mu_i \sum_{p=1}^{n_\mu} \chi^{\mu,A_p}_i \sim r^2 \xi^\mu_i (1+\sqrt \rho_{\mu} z^{\mu}_i) $, with $z^\mu_i$ normally distributed and $\rho_\mu = (1-r^2)/(n_\mu r^2)$. Thus, in this regime, we get $\text{sgn}(\sum_{p=1}^n \xi_i^{\mu_p,A_p})\approx\text{sgn}(\sum_{\mu=1}^n \xi_i^{\mu})$, resulting again in a spurious combination of patterns. We use configurations of the form $\bb x^\gamma$ (where now $\gamma$ labels the $m$ different realizations of the mini-batch) as input configurations for the model in (\ref{eq:energy}) and reconstruct patterns with the same procedure as before. Our findings are reported in Fig.~\ref{fig:results_2}. Again, high-complexity machines have better extraction capabilities.
Notably, in all situations the extraction procedure appears to be very robust w.r.t. to intrinsic noise in the dataset (even for high values of the mini-batch entropy $\rho$), as clearly shown by the weak dependence on $r$ of the fraction $K_R/K$. In fact, as explained above, employing combinations of data points filters out the intrinsic noise, with these states being -- at finite $r$ -- almost indistinguishable from usual spurious configurations of patterns. Therefore, the machine is expected to work nicely for the task under consideration.

\begin{figure}[h!]
    \centering
    \includegraphics[width=0.8\textwidth]{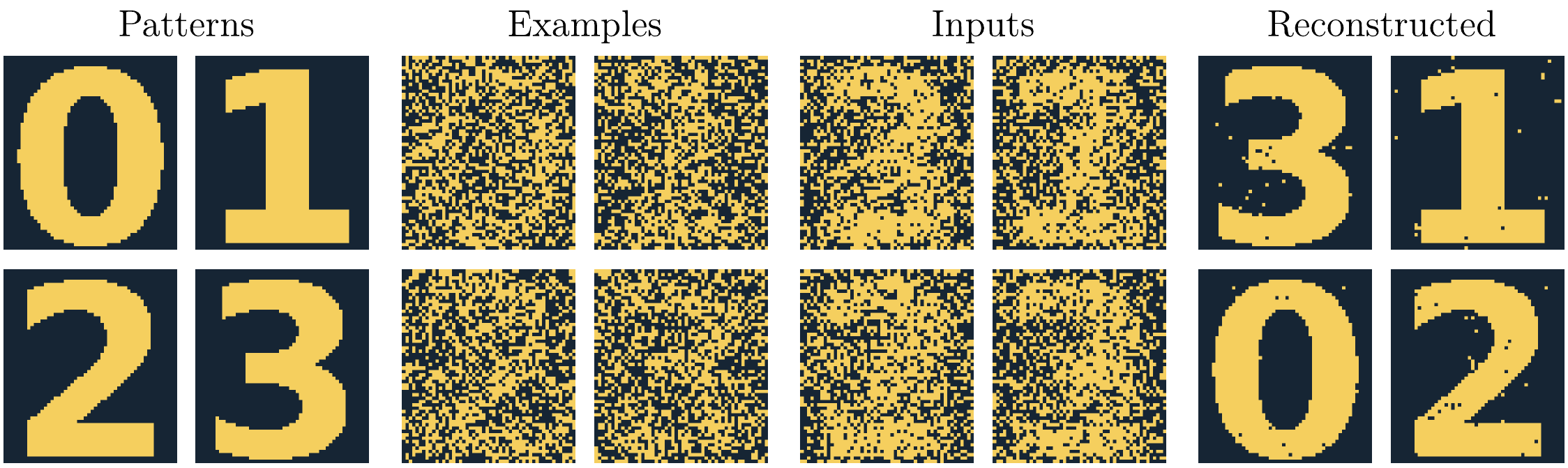}
    \caption{Summary of results for the pattern reconstruction by unsupervised structured examples. In the left block, we report the hidden patterns we want to reconstruct, starting from a very noisy dataset ($r=0.2$) a sample of which is presented in the second block from the left. The number of examples per class is $M=5000$, from which we generate $m=50$ different mini-batches of size $n=10$, which are used to generate the input configurations. In the right column, we reported the results of the pattern reconstruction. The network size is $N=3016$ (images have size $58\times52$), the parameters are $\beta=4$, $\lambda=0.2$, $H=0.05$ and $L=4$.}
    \label{fig:results_3}
\end{figure}

As a last experiment, we test the procedure on a structured (but still simple) dataset. We take as patterns a synthetic realizations of the first 4 digits, we realize the dataset again with multiplicative noise, and consider vectors $x_i ^\gamma =\text{sgn}(\sum_{p=1}^n \xi_i^{\mu_p,A_p})$ built by $m$ mini-batches of size $n$. Then, we perform the pattern extraction procedure.\footnote{Since, in the structured dataset, intrinsic features would have a higher mutual correlation w.r.t. the random case, we relax the eligibility condition of final configurations by considering duplicates two states with mutual overlap $q_{lk}>0.9$. The ``almost eigenvectors'' criterion for the Kohonen kernel is left unchanged.} As we have shown in the previous experiment, the pattern reconstruction procedure is robust against data noise. In the case under consideration, the dataset is indeed generated with very poor quality ($r=0.2$). The final results are reported in Fig. \ref{fig:results_3}. Even starting with visually unrecognizable samples, taking spurious combinations of examples filters out the noise, so that the system is able to effectively reconstruct the hidden patterns. The average quality of overlap between the reconstructed patterns and the true ones is very high, that is $\langle N^{-1} \bb\xi_R \cdot \bb \xi\rangle \approx 0.98$.
\section{Conclusions}
We presented a procedure to reconstruct hidden patterns starting from partial information, namely Hebb's coupling matrix and additional information in terms of spurious combinations of the patterns. We extensively used the $L$-direction associative memories, allowing for a parallel retrieval of the patterns by disentangling such spurious states. We analyze the procedure in three settings, namely random patterns, synthetic and structured noisy datasets, always leading to high-quality reconstruction of the hidden features. We intend to deepen the results here reported in order to extend the possibility to known higher-order spatial moments of the patterns by suitably modifying the energy function (for instance, adding dense contributions) as well as hyper-parameter fine-tuning (possibly by means of a statistical-mechanical approach), and applying the procedure to realistic datasets.

\subsubsection*{Acknowledgments}
EA and AF acknowledge financial support from PNRR MUR project PE0000013-FAIR and
from Sapienza University of Rome (RM120172B8066CB0, AR2221815D7192C1,
AR1221815EA97525). AA is member of GNFM-INdAM, which is acknowledged.

\bibliography{nfam2025_workshop}

\begin{thebibliography}{10}

\bibitem{hopfield1982neural}
JJ~Hopfield.
\newblock Neural networks and physical systems with emergent collective
  computational abilities.
\newblock {\em Proceedings of the national academy of sciences},
  79(8):2554--2558, 1982.

\bibitem{amit1987statistical}
DJ~Amit, H~Gutfreund, and H~Sompolinsky.
\newblock Statistical mechanics of neural networks near saturation.
\newblock {\em Annals of physics}, 173(1):30--67, 1987.

\bibitem{dotsenko1991statistical}
VS~Dotsenko, ND~Yarunin, and EA~Dorotheyev.
\newblock Statistical mechanics of {H}opfield-like neural networks with
  modified interactions.
\newblock {\em Journal of Physics A: Mathematical and General}, 24(10):2419,
  1991.

\bibitem{fachechi2019dreaming}
A~Fachechi, E~Agliari, and A~Barra.
\newblock Dreaming neural networks: forgetting spurious memories and
  reinforcing pure ones.
\newblock {\em Neural Networks}, 112:24--40, 2019.

\bibitem{krotov2016dense}
D~Krotov and JJ~Hopfield.
\newblock Dense associative memory for pattern recognition.
\newblock {\em Advances in neural information processing systems}, 29, 2016.

\bibitem{camilli2023matrix}
F~Camilli and M~M{\'e}zard.
\newblock Matrix factorization with neural networks.
\newblock {\em Physical Review E}, 107(6):064308, 2023.

\bibitem{bengio2012unsupervised}
Y~Bengio, AC~Courville, and P~Vincent.
\newblock Unsupervised feature learning and deep learning: A review and new
  perspectives.
\newblock {\em CoRR, abs/1206.5538}, 1(2665):2012, 2012.

\bibitem{aiudi2025local}
R~Aiudi, R~Pacelli, P~Baglioni, A~Vezzani, R~Burioni, and P~Rotondo.
\newblock Local kernel renormalization as a mechanism for feature learning in
  overparametrized convolutional neural networks.
\newblock {\em Nature Communications}, 16(1):568, 2025.

\bibitem{negri2023storage}
M~Negri, C~Lauditi, G~Perugini, C~Lucibello, and E~Malatesta.
\newblock Storage and learning phase transitions in the random-features
  {H}opfield model.
\newblock {\em Physical Review Letters}, 131(25):257301, 2023.

\bibitem{negri}
S~Kalaj, C~Lauditi, G~Perugini, C~Lucibello, EM~Malatesta, and M~Negri.
\newblock Random {F}eatures {H}opfield {N}etworks generalize retrieval to
  previously unseen examples.
\newblock {\em arXiv preprint arXiv:2407.05658}, 2024.

\bibitem{agliari2024regularization}
E~Agliari, F~Alemanno, M~Aquaro, and A~Fachechi.
\newblock Regularization, early-stopping and dreaming: a {H}opfield-like setup
  to address generalization and overfitting.
\newblock {\em Neural Networks}, 177:106389, 2024.

\bibitem{agliari2025networks}
E~Agliari, A~Alessandrelli, A~Barra, MS~Centonze, and F~Ricci-Tersenghi.
\newblock Networks of neural networks: more is different.
\newblock {\em arXiv preprint arXiv:2501.16789}, 2025.

\bibitem{kohonen1973representation}
T~Kohonen and M~Ruohonen.
\newblock Representation of associated data by matrix operators.
\newblock {\em IEEE Transactions on Computers}, 100(7):701--702, 1973.

\bibitem{personnaz1985information}
L~Personnaz, I~Guyon, and G~Dreyfus.
\newblock Information storage and retrieval in spin-glass like neural networks.
\newblock {\em Journal de Physique Lettres}, 46(8):359--365, 1985.

\bibitem{kanter1987associative}
I~Kanter and Haim Sompolinsky.
\newblock Associative recall of memory without errors.
\newblock {\em Physical Review A}, 35(1):380, 1987.

\end{thebibliography}
\bibliographystyle{unsrt}

\appendix
\section{Details on numerical computations}
Experiments are conducted by initializing each layer with a generic spurious observation $\bb x^\gamma$, and then evolving the system according to the dynamics described in Eq.~\ref{eq:update}, using a parallel update scheme (i.e., all neurons across the entire network are updated simultaneously). The dynamics are run for a sufficiently long time to ensure thermalization toward a fixed point. Unless otherwise specified, the total number of parallel updates is set to 5000. Numerical simulations were performed using TensorFlow 2.11 with CUDA Toolkit 11.7 and cuDNN 8.5, on an NVIDIA GeForce RTX 4070 Ti GPU.

\section{Sensitivity to hyperparameters on reconstruction performances}

\begin{figure}[h!]
    \centering
    \includegraphics[width=\textwidth]{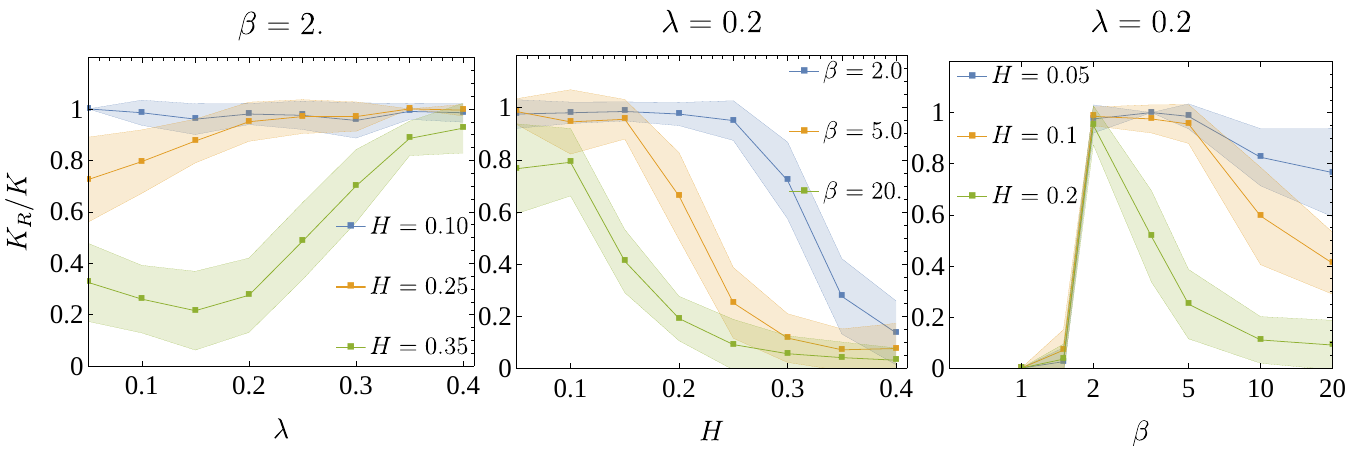}
    \caption{Sensitivity of the model's reconstruction capabilities to hyperparameters.
In the three plots, we explore sections of the hyperparameter space by computing the fraction of reconstructed patterns, $K_R/K$, while fixing one hyperparameter and varying the other two. In the left plot, we fix $\beta = 2$ and analyze the dependence of $K_R/K$ on $\lambda$ for various values of $H$. In the center plot, we fix $\lambda = 0.2$ and examine how the reconstruction performance varies with $H$ for different values of $\beta$. Finally, in the right plot, we report the dependence of $K_R/K$ on $\beta$, fixing $\lambda = 0.2$ and varying the external field $H$. The shaded regions represent intervals of width two standard deviations centered around the mean. Results are averaged over 20 independent realizations of the patterns. The network size is $N = 1000$, the number of patterns is $K = 10$, and the number of layers is $L = 3$.}
    \label{fig:results_hps}
\end{figure}

In this appendix, we explore how the model’s reconstruction capabilities depend on the control parameters. We take a numerical approach, as a full theoretical understanding of the reconstruction regimes across the hyperparameter space would require a statistical mechanical analysis--this lies beyond the scope of the present work. 
For simplicity, we focus on the first setting, where the available information consists of spurious combinations of patterns, and the control parameters are $\beta$, $\lambda$, and $H$. To reduce the computational cost of exploring a three-dimensional hyperparameter space, we analyze two-dimensional sections by fixing one hyperparameter and varying the other two over a range of reasonable values. The results of this analysis are presented in Fig.~\ref{fig:results_hps}. First, note that successful disentanglement of spurious pattern combinations requires the temperature to be not too high -- thus avoiding an ergodic behavior --but still sufficiently high to allow the model to explore the energy minima landscape. We start by fixing $\beta=2$ and vary $\lambda$ and $H$. In the left plot, we see that, for the given level of thermal noise, the behavior of the reconstruction capabilities in $\lambda$ is crucially dependent on $H$. In particular, for a low external field ($H = 0.1$), good reconstruction is achieved across a broad range of $\lambda$ values ($\lambda = 0.05 \div 0.4$). This suggests that at $\beta = 2$ and $H = 0.1$ the model is relatively robust to variations in $\lambda$. A similar analysis can be carried out by fixing $\lambda = 0.2$ and varying $H$ across different values of $\beta$. As $\beta$ increases, the range of $H$ values that yield good reconstruction performance becomes narrower and shifts toward lower values. This observation further supports the choice of $\beta = 2$ as a balanced setting for effective reconstruction. Finally, in the right plot, we perform a consistency check on the choice of temperature by fixing $\lambda = 0.2$ and varying both $H$ and $\beta$. In all cases, the highest reconstruction performance is observed at $\beta = 2$. 
Our chosen setting — $\beta = 2$, $\lambda = 0.2$, and $H = 0.1$ — lies well within this favorable region. Naturally, a similar type of analysis can be carried out in the case of noisy realizations of structured patterns, which motivates the different parameter choices adopted in the third experiment.


\section{The acceptance criterion}
\begin{figure}[ht!]
    \centering
    \includegraphics[width=\textwidth]{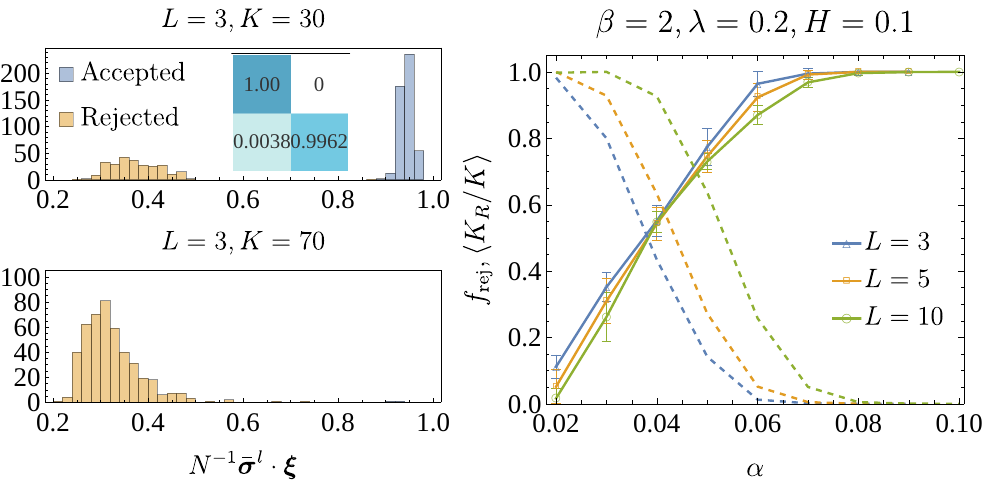}
    \caption{Effectiveness of the acceptance criterion. In the left column, we compare the fractions of the accepted final configurations (blue histogram) w.r.t. the discarded ones (yellow histogram) as a function of their overlap with the hidden patterns. For $L=3$ and low $K=30$ (upper left plot), the acceptance criterion is able to distinguish between reconstructed truths and their spurious combinations, and the effectiveness is high (see the confusion matrix in the inset plot). For higher values of $K$ (upper right), the thermalization of the systems more likely ends up in spurious configurations, which are rejected in bulk, resulting in a loss of reconstruction power. In the right plot, we report the fraction of rejected configurations as a function of $\alpha$ for $L=3$ (blue), $5$ (yellow) and $10$ (green). For the sake of completeness, in dashed lines we also reported the associated results for the average fraction of reconstructed patterns. The size of the network is $N=1000$, the parameters are $\beta=2$, $\lambda=0.2$ and $H=0.1$, the number of spurious observation is $m=50$. Results are averaged over 20 different realizations of the hidden patterns.}
    \label{fig:results_KS}
\end{figure}

As previously mentioned in the main text, the acceptance criterion for a reconstructed pattern involves a two-step verification process. First, we ensure that the final configurations of each layer exhibit low mutual overlap. This step eliminates potential duplicates in the final sample. Second, we verify that $\bar{\bb\sigma}^l \bb J^{K} \bar{\bb\sigma}^l / N > 0.8$, where $\bb J^{K}$ denotes the pseudo-inverse coupling matrix. This condition serves to filter out failed reconstructions resulting from relaxation towards spurious states. In this appendix, we further elaborate the effectiveness of the second step. Indeed, for any pattern $\bb \xi^\mu$, we have that
$$
\sum_{j=1}^N J^{K}_{i,j}\xi^\mu_j = \sum_{j=1}^N \frac1N \sum_{\nu,\rho=1}^K \xi^\nu_i C^{-1}_{\nu,\rho}\xi^\rho_j \xi^\mu _j= \sum_{\nu,\rho=1}^K \xi^\nu_i C^{-1}_{\nu,\rho}C_{\rho,\mu}=\sum_{\nu=1}^K \xi^\nu_i \delta_{\nu\mu}=\xi^\mu_i.
$$
Thus, the eigenspace associated with the eigenvalue 1 of the pseudo-inverse coupling matrix is $K$-dimensional and consists solely of linear combinations of the true patterns. Spurious states are thus excluded from this eigenspace due to the non-linearity of the sign function. Furthermore, by multiplying both sides of the equation by $\xi_i^\mu$ and summing over the index $i$, we have
$$
\sum_{i,j=1}^N\xi^\mu_i J^{K}_{i,j}\xi^\mu_j =\sum_{i=1}^N (\xi^\mu_i )^2= N.
$$
Therefore, the condition $\bar{\bb\sigma}^l \bb J^K \bar{\bb\sigma}^l / N = 1$ would ideally fulfill the desired acceptance criterion. However, in practice, this is rarely achieved due to two main reasons: i) the candidate configurations $\bar{\bb\sigma}^l$ are, at best, stochastic realizations of the underlying patterns, meaning that a finite fraction of bits may be misaligned with the corresponding true pattern; and ii) the pseudo-inverse matrix $\bb J^K$ is itself obtained through an iterative algorithm, which may introduce numerical approximations or deviations from the exact theoretical construction. Thus, we need to relax the acceptance criterion allowing for states with $\bar {\bb \sigma}^l \bb J^K \bar {\bb \sigma}^l /N$ above a sufficiently high threshold. Here, this threshold is fixed to $0.8$. In Fig. \ref{fig:results_KS} we give numerical results supporting the validity of our criterion.

\par\medskip

In the left column, we display histograms of the overlap $\frac{1}{N} \bar{\bb\sigma}^l \cdot \bb\xi$ between the candidate configurations and the hidden patterns. Specifically, the blue histogram corresponds to configurations that satisfy the acceptance criterion, while the yellow histogram represents those that violate the condition $\bar{\bb\sigma}^l \bb J^K \bar{\bb\sigma}^l / N > 0.8$. As is clear, this criterion generally succeeds in filtering out states that result from the system thermalizing into spurious combinations of the patterns. For sufficiently low values of $K$, a fraction of the configurations $\bar{\bb\sigma}^l$ satisfy the acceptance criterion, and all of these exhibit a high overlap with the hidden patterns. In contrast, the rejected configurations typically show an overlap $\frac{1}{N} \bar{\bb\sigma}^l \cdot \bb\xi \le 0.5$, consistently with the expectation that they correspond to spurious pattern combinations. In the inset, we also report a normalized confusion matrix supporting the validity of the criterion. The structure of this matrix is the following:
$$
\bb\Gamma = \begin{pmatrix}
    \frac{TP}{TP+FP} & \frac{FP}{TP+FP}\\
    \frac{FN}{TN+FN} & \frac{TN}{TN+FN}
\end{pmatrix},
$$
where true positives (TP) refer to configurations $\bar{\bb\sigma}^l$ that satisfy the acceptance criterion and exhibit a high overlap with the patterns (e.g., $\frac{1}{N} \bar{\bb\sigma}^l \cdot \bb\xi \ge 0.8$). False positives (FP) are those configurations that are accepted by the criterion but have low correlation with the ground-truth patterns (i.e., $\frac{1}{N} \bar{\bb\sigma}^l \cdot \bb\xi < 0.8$). Conversely, true negatives (TN) are configurations rejected by the criterion that indeed show low overlap, while false negatives (FN) are those that are incorrectly rejected despite exhibiting high overlap with the patterns. Although these FN cases are discarded, they do not significantly affect the overall reconstruction performance of the model. Since the fraction of true positive and true negative states is close to 1, we conclude that the acceptance criterion effectively distinguishes between accurate reconstructions and spurious combinations of the hidden patterns. As expected, increasing $K$ leads to a larger fraction of states failing the sanity check: in this regime, pattern retrieval becomes significantly more challenging, and the reconstruction process tends to break down. This behavior is illustrated in the right-hand plot, where we show the fraction of rejected configurations as a function of $\alpha = K/N$ for $L = 3, 5, 10$, along with the corresponding average reconstruction performance $K_R/K$ (shown as dashed lines). As the information load $\alpha$ increases, the likelihood that the system thermalizes into spurious states also grows, compromising the model’s reconstruction accuracy. However, increasing the number of layers $L$ improves the acceptance rate, thereby enhancing the ability to retrieve patterns even under higher storage demands. Investigating the optimal scaling relations between the hyperparameters, the number of layers, and the storage capacity is a crucial aspect of this framework. However, a thorough analysis of this problem within a statistical mechanical perspective is beyond the scope of the present work and will be addressed in future studies.

\section{Finite-size scaling}
\begin{figure}[tb]
	\centering
	\includegraphics[width=0.65\textwidth]{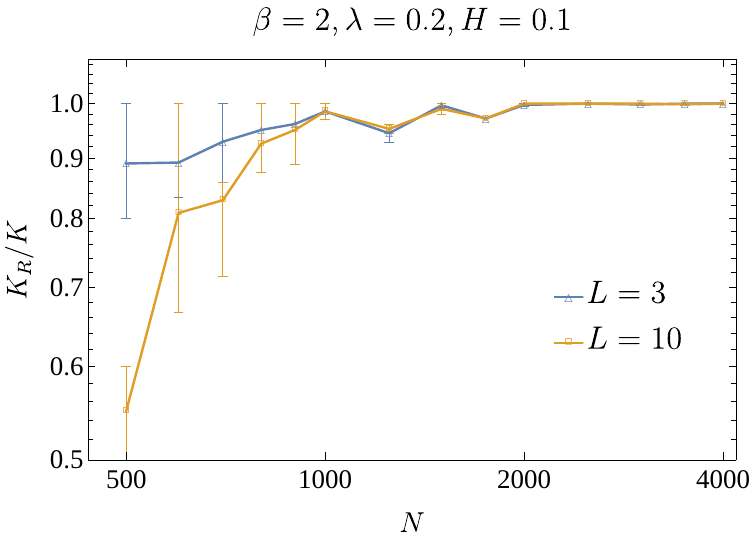}
	\caption{Finite-size scaling w.r.t. the layer size. The plot show the results of reconstruction capabilities for varying layer size $N$ and $L=3,10$. Results are averaged over 20 different realizations of the patterns. The model parameters are $\beta=2$, $\lambda=0.2$, $H=0.1$. The number of observation is fixed to $m=2m_{\text{min}} (K)$. The number of dynamics updates of each network is fixed to $5\cdot N$.}
	\label{fig:results_FSS}
\end{figure}
As a final point, we examine the robustness of the model's reconstruction capabilities with respect to the individual layer size $N$. To ensure a fair comparison, networks of different sizes must operate under equivalent conditions. First, the number of stored patterns should scale with $N$, i.e., $K = \alpha N$. However, increasing $K$ while keeping $m$ fixed significantly reduces the probability of successfully reconstructing all patterns; in other words, also $m$ should scale with $N$. To estimate this scaling, we considered a related problem. Suppose we have a collection of $K$ objects, from which we uniformly sample a subset of $L$ elements in each experiment (i.e., each object is selected with probability $1/K$). We repeat this experiment $m$ times, replacing the extracted elements after each trial. Our goal is to compute the probability that all $K$ patterns are observed at least once across the $m$ trials. Consider a fixed element, say $\mu = 1$. The probability that it is not selected in a single trial is approximately $(1 - 1/K)^L \approx 1 - L/K$, assuming $K$ is large. Therefore, the probability that this element is never observed over $m$ independent repetitions is $(1-L/K)^m$. From this, we can say that, for $K$ large enough, the probability that at least one of the $K$ elements is never observed across all trials is approximately $\approx K(1-L/K)^m$. Since this is the complementary event to the one we are interested in, we can conclude that the probability of extracting all of the patterns at least once is approximately
$$
P(\bb \xi^1,\dots,\bb\xi^K \text{ observed}) =1-K \Big(1-\frac L K\Big)^m.
$$
This represents an ideal scenario for our setting, in which each layer extracts exactly $L$ distinct patterns at each step, without generating duplicates or failing to reconstruct any ground-truth. To ensure a high probability of observing all $K$ patterns, we impose the condition $P(\bb \xi^1,\dots,\bb\xi^K \text{ observed})=1-\epsilon$ with $\epsilon$ being the tolerance against failed experiments. Thus, we can thus set ${\textstyle K (1- L/ K)^m=\epsilon}$ so that, expanding at the leading contribution in $K$, we get
$$
m_{\text{min}}(K)\approx \frac KL \log \frac {K}\epsilon.
$$

\par\medskip
In our experiments, we fix $\alpha=0.01$, $\epsilon=0.01$ and $m=2m_{min}(K)$. The results of the finite-size scaling analysis are reported in Fig. \ref{fig:results_FSS} for $\beta=2$, $\lambda =0.2$ and $H=0.1$, with $L=3,10$. As evident from the plot, apart from the lower performance observed at small $N$ -- which lies outside the regime where the scaling approximation holds -- the reconstruction capabilities remain consistently high. Moreover, they are robust with respect to both the layer size $N$ and the number of layers $L$.

\end{document}